\documentstyle[preprint,aps]{revtex}

\begin{document}
\draft
\title{Exhibition of the periodicity of Quantum Fourier Transformation in Nuclear
Magnetic Resonance}
\author{Xinhua Peng$^{1}$, Xiwen Zhu$^{1\thanks{%
Corresponding author. {E-mail:xwzhu@wipm.whcnc.ac.cn; Fax: 0086-27-87885291.}%
}}$, Ximing Fang$^{2,1}$, Mang Feng$^{1}$, Xiaodong Yang$^{1}$, Maili Liu$%
^{1},$ and Kelin Gao$^{1}$}
\address{$^{1}$Laboratory of Magnetic Resonance and Molecular Physics, \\
Wuhan Institute of Physics and Mathematics, The Chinese Academy of Sciences,%
\\
Wuhan, 430071, People's Republic of China\\
$^{2}$Department of Physics, Hunan Normal University, Changsha, 410081,\\
People's Republic of China}
\maketitle

\begin{abstract}
The remarkable capability of quantum Fourier transformation (QFT) to extract
the periodicity of a given periodic function has been exhibited by using
nuclear magnetic resonance (NMR) techniques. Two separate sets of
experiments were performed. In a full QFT, the periodicity were validated
with state tomography and fidelity measurements. For a simplified QFT, the
three-qubit pseudo-pure state was created by introducting an additional
observer spin, and the spectra recorded on the observer spin showed
intuitively the power of QFT\ to find the periodicity. Experimentally
realizing the QFT provides a critical step to implement the renowned Shor's
quantum factoring algorithm and many other algorithms. Moveover, it can be
applied to the study of quantum chaos and other quantum information
processing.
\end{abstract}



The most striking discovery in quantum computation is that quantum computers
can efficiently perform some tasks which are not feasible on a classical
computer. For example, the most famous of the current quantum algorithms ---
Shor's quantum factoring algorithm\cite{Shor}, can factor a number
exponentially faster than the best known classical algorithms, which leads
to the cracking the RSA encryption system\cite{Rivest}. However, the key to
Shor's algorithm rests in an essential way on the application of a suitable
quantum Fourier transform to efficiently determine the periodicity of a
given periodic function\cite{Nielsenbook}. The QFT is the key ingredient for
not only Shor's algorithm but also many other interesting quantum
algorithms, including Simon's algorithm (Simon 1994)\cite{Simon,Jozsa},
Kitaev's algorithm\cite{Kitaev,Abrams}, the estimating arbitrary phase
algorithm\cite{Kitaev,cleve} and the ordering-finding algorithm\cite{cleve}
etc.. Moveover, the scheme of exploiting quantum computation to study
quantum chaos proposed by Schack\cite{Schack}, Brun and Schack\cite{Brun},
are also based on the QFT.

For an arbitrary positive integer $x$, the QFT can be defined by a unitary
transformation\cite{Coppersmith,Ekert,Nielsenbook}

\begin{equation}
QFT:|x\rangle \rightarrow \frac{1}{\sqrt{N}}\sum_{y=0}^{N-1}e^{2\pi
ixy/N}|y\rangle ,\qquad for\text{ }x\in \left\{ 0,1,2,...,2^{n}-1\right\} ,
\end{equation}
where $N=2^{n},$ with $n$ being the qubit number. Recently the QFT have been
realized on a three bit NMR quantum computer. However, the implementation of
the QFT was performed on the thermal equilibrium state\cite{QFTNMR} so that
the striking power of the extraction of a periodicity wasn't exhibited from
the experimental results.

Imagine an oracle that computes a function $f:Z_{n}\rightarrow Z_{m}$ that
has an unknown period $r$ $(1\leq r\leq 2^{n})$:

\begin{equation}
f(x+mr)=f(x),
\end{equation}
where $m$ is any integer such that $x$ and $x+mr$ lie in $\left\{
0,1,2,...,2^{n}-1\right\} .$ The goal is to find the period $r$ of $f(x)$.
Classically, this problem is hard, requiring $O\left( 2^{n}\right) $ queries
to hit two equal values with high probability\cite{Maslen}. But a quantum
algorithm described as follows can find $r$ in time poly$(n)$\cite
{Ekert,Ekert1}.

Define a unitary transform $U$: 
\begin{equation}
U|x\rangle |y\rangle \rightarrow |x\rangle |y\oplus f(x)\rangle ,
\end{equation}
where $\oplus $ denotes addition modulo 2. One first initialize two
registers to $\left| 0\right\rangle \left| 0\right\rangle ,$ then create an
equal superposition $\frac{1}{\sqrt{2^{n}}}\sum_{x}\left| x\right\rangle $
in the first register (easily prepared by applying $H^{\left( n\right) }).$
Owing to quantum parallelism, querying the oracle and applying $U$ prepare
the state 
\begin{equation}
\left| f\right\rangle =\frac{1}{\sqrt{N}}\sum_{x=0}^{N-1}\left|
x\right\rangle \left| f\left( x\right) \right\rangle .
\end{equation}
By measuring the second register, the result $\left| f\left( x_{0}\right)
\right\rangle $ for some $0\leq x_{0}\leq r-1$ is obtained .In the meantime,
the first register is collapsed into the state 
\begin{equation}
\left| \psi \right\rangle =\frac{1}{\sqrt{K}}\sum_{j=0}^{K-1}\left|
x_{0}+jr\right\rangle ,
\end{equation}
where $N-r\leq x_{0}+(K-1)r\leq N.$ The periodicity $r$ is not explicitly
exhibited due to the offset $x_{0}.$

Applying QFT to the state $\left| \psi \right\rangle ,$ one gets 
\begin{equation}
F\left| \psi \right\rangle =\frac{1}{\sqrt{r}}\sum_{j=0}^{r-1}e^{2\pi i\frac{%
x_{0}j}{r}}\left| j\frac{N}{r}\right\rangle .
\end{equation}
A value $c$ is supposed to be obtained by measuring the register, where $c$
is necessarily a multiple of $N/r$, i.e., $c/N=\lambda /r,(0\leq \lambda
<r). $ If $\lambda $ is fortuitously coprime to $r$, $r$ can be determined
by cancelling $c/N$ down to an irreducible fraction. According to the prime
number theorem\cite{Ekert,Hardy,Schroeder}, the probability that $\lambda $
is coprime to $r$ is at least $1/logr$ which exceeds $1/logN$. Hence
repeating the above procedure $O(logN)$ times one can succeed in determining 
$r$ with any prescribed probability $1-\epsilon $ as close to $1$ as desired.

Another way of analyzing the distribution of outcomes obtained is to
calculate the reduced density matrix for the first register and calculate
the measurement statistics. As the NMR quantum computer is ensemble quantum
computation, we can achieve the statistical results of the probability
distribution of the final state. It can be seen from Eq. (5) and (6) that,
the QFT transforms the input state with the period $r$ into the output state
with the period $k=N/r$ containing $r$ items due to the displacement
invariance of the QFT. For example, if the 3-qubit state $\left| \psi
\right\rangle $ in Eq. (5) input to QFT is of the form 
\begin{equation}
\left| \psi \right\rangle =\frac{1}{2}(\left| 0\right\rangle +\left|
2\right\rangle +\left| 4\right\rangle +\left| 6\right\rangle )
\end{equation}
or 
\begin{equation}
\left| \psi \right\rangle =\frac{1}{2}(\left| 1\right\rangle +\left|
3\right\rangle +\left| 5\right\rangle +\left| 7\right\rangle ),
\end{equation}
the state periodicity should be $r=2$. Applying the QFT, one gets 
\begin{equation}
QFT:\left| \psi \right\rangle =\frac{1}{\sqrt{2}}(\left| 0\right\rangle
+\left| 4\right\rangle )
\end{equation}
or 
\begin{equation}
QFT:\left| \psi \right\rangle =\frac{1}{\sqrt{2}}(\left| 0\right\rangle
-\left| 4\right\rangle ).
\end{equation}
From the state periodicity after the QFT $k=4$ one can infer the state
periodicity before the QFT $r=N/k=2$, which is consistent with the state Eq.
(7) or (8). Contrarily, if $\left| \psi \right\rangle $ has the form of Eq.
(9) or (10), the state periodicity after the QFT $k=2$ so that $r=N/k=4$.

We have demonstrated these effects by liquid-state NMR using the carbon-13
labeled alanine $NH_{3}^{+}-C^{\alpha }H(C^{\beta }H_{2})-C^{^{\prime
}}O_{2}^{-}$ dissolved in $D_{2}O.$ All NMR experiments were performed on a
Bruker ARX500 spectrometer with respect to transmitter frequencies of 500.13
MHz ($^{1}H$) and 125.77 MHz ($^{13}C$). The measured NMR parameters are
listed in Table 1. We can see from above that, the second register is only
used to prepare the periodic state in the first register. Thus, in actual
experiments, we just used the first register for determining the
periodicity. In addition, owing to the nature of NMR ensemble, the display
of the final state were performed by two different methods, tomography\cite
{Chuang} and ''spectral implementation''\cite{Madi}.

The logic network for the QFT is shown in Fig. 1\cite
{cleve,Coppersmith,Beckman}, consisting of a one-qubit gate 
\begin{equation}
H_{j}=\frac{1}{\sqrt{2}}\left( 
\begin{array}{ll}
1 & 1 \\ 
1 & -1
\end{array}
\right) ,
\end{equation}
acting on qubit $j$ and a controlled-$R_{d}$ gate 
\begin{equation}
R_{d}=\left( 
\begin{array}{cc}
1 & 0 \\ 
0 & e^{i\pi /2^{d}}
\end{array}
\right)
\end{equation}
acting on qubit $j$ conditional on qubit $k$ being in the $\left|
1\right\rangle $ state with $d=j-k.$ Here the $H_{j}$ gate can be realized
by the NMR pulse sequence $X_{j}(\pi )Y_{j}\left( \frac{\pi }{2}\right) $ or 
$Y_{j}\left( -\frac{\pi }{2}\right) X_{j}(\pi ),$ and the controlled-$R_{j}$
gate, by the pulse sequence $Z_{j}(\frac{\pi }{2})Z_{k}\left( \frac{\pi }{2}%
\right) J_{jk}\left( -\frac{\pi }{2}\right) $\cite{Gershenfeld,Jones}. By
carefully choosing the expressions for the Hadamard and $Z$ gates and
exploiting the commutation relations\cite{Ernst} to eliminate as many
unnecessary operations as possible, the complete pulse for the QFT can be
reduced as: 
\begin{eqnarray}
&&X_{1}(-\frac{5\pi }{8})Y_{1}\left( \frac{\pi }{2}\right) J_{21}\left( -%
\frac{\pi }{2}\right) J_{31}\left( -\frac{\pi }{4}\right) X_{2}(-\frac{\pi }{%
2})Y_{2}\left( -\frac{\pi }{4}\right)  \nonumber \\
&&X_{2}\left( -\frac{\pi }{4}\right) Y_{2}\left( \frac{\pi }{2}\right)
J_{32}\left( -\frac{\pi }{2}\right) Y_{3}\left( -\frac{\pi }{2}\right)
X_{3}(-\frac{5\pi }{8})
\end{eqnarray}
(applied from the left to the right). Finally a swap operation $S_{13}$ are
used to reverse the order of qubits to obtain the desired output from the
QFT. We could just as easily relabel qubits 1, 2 and 3 in place of the Swap
gate. Thus, we need not perform an actual physical swap gate here; a mental
relabeling of the qubits is sufficient.

We preformed two separate sets of experiments. In the first set, the full
QFT was executed on a three-qubit quantum computer. Three $^{13}C$ nuclei
were chosen as three qubits and protons were decoupled during the whole
experiments by using a standard heteronuclear decoupling technique. All $%
^{13}C$ nuclei were taken 0.7ms for selective $\frac{\pi }{2}$ pulses of a 
{\it Gaussian} shape. At the beginning of the experiment, we labeled $%
C^{\prime },$ $C^{\alpha }$ and $C^{\beta }$ as spin 1, 2 and 3,
respectively. After the QFT, we identify spin 1 with $C^{\beta }$ spin 3
with $C^{^{\prime }}.$

The pseudo-pure state was prepared from the thermal equilibrium state by the
procedure summarized in Table 2, in which the magnetic field gradients
(denoted by $G_{z}$ ) to dephase off-diagonal elements of the density matrix
at strategic points along the way were used\cite{Ernst}. Owing to the
difference of the relaxation times between nuclei, we used a $\pi $ rf pulse
and a spell (the reversal recovery) to balance the effect of the different
relaxation. Meanwhile step (1) in Table 2 was replaced by the spell. By
adjusting carefully the spell, the experimental spectra for a pseudo-pure
state $\rho _{000}=I_{1}^{\alpha }I_{2}^{\alpha }I_{3}^{\alpha }$ ($%
I_{i}^{\alpha }=\left| 0\right\rangle \left\langle 0\right| =\frac{1}{2}%
+I_{iz})$ were finally recorded (shown in Fig. 2b) through reading-out
pulses on the spectrometer. The normalized deviation density matrix $\rho
_{000}$ was confirmed by full tomography\cite{Chuang} shown in Fig. 3a.

The state in Eq. (7) and Eq. (9) can be prepared by the Hadamard gates $%
H_{1}H_{2}$ and $H_{1}$ from the pseudo-pure state, respectively. Applying
the sequence (13), the experimental spectra were shown in Fig. 2 (c) and
(d). The density matrices were reconstructed from the obtained experimental
spectra. The real parts of these matrices are shown in Fig. 3 (b) and (c),
respectively. The fidelity of the QFT were calculated using the measure,
called the {\it attenuated correlation}\cite{Teklemariam} 
\begin{equation}
c(\hat{\rho}^{\exp })=\frac{Tr(\hat{\rho}^{th}\hat{\rho}^{\exp })}{Tr(\hat{%
\rho}^{th}\hat{\rho}^{th})}.
\end{equation}
Here, $\hat{\rho}^{th}$ is defined as a theoretical output state,
transformed by the ideal transformation on a computer for the measured
pseudo-pure state $\hat{\rho}_{000}^{\exp }$, and an experimentally
implemented control sequence for the same transformation on the spectrometer
to get $\hat{\rho}^{\exp }.$ The values of the correlation for each of the
three tomographic readouts were $c(\hat{\rho}_{000}^{\exp })=1$ (by
definition), $c(\hat{\rho}_{qft1}^{\exp })=0.92$ and $c(\hat{\rho}%
_{qft2}^{\exp })=0.84.$ The measure reflects the imperfections of the
experiments, including inhomogeneity of RF fields and static magnetic fields
and imperfect calibration of rotations, gradient pulses and relaxation.

In the second set of experiments, we prepared a pseudo-pure state with
introduction of an observer spin and employed a further simplification of
the QFT proposed by Preskill\cite{Preskill}, in which no two-qubit gates are
needed at all and only $n$ Hadamard gates and $n-1$ single-qubit rotations
are applied. Its key thoughts is the symmetry of the controlled-$R_{d}$ gate
on the two qubits, and to measure a single qubit first and then apply the
controlled-$R_{d}$ gate to the next qubit, conditioned on the outcome of the
measurement of the qubit, instead of applying controlled-$R_{d}$ and then
measuring. Of course, it is only suitable for a special state, for example,
the computational basis state $\left| x_{3}\right\rangle \left|
x_{2}\right\rangle \left| x_{1}\right\rangle ,$ ($x_{i}=0$ or $1$) and the
state in Eq. (7) and (9).

The pseudo-pure state $I_{0z}I_{1}^{\alpha }I_{2}^{\alpha }I_{3}^{\alpha }$
with introduction of an observer spin $I_{0}$ were prepared by a spatial
averaging method proposed by Sakaguchi et al.\cite{Sakaguchi}, the procedure
shown in Table 3. We chose $C^{\alpha }$ as the obssever spin $I_{0}$ due to
resolved scalar $J$ couplings to all other spins and $C^{\prime },$ $%
C^{\beta }$ and $H$ being joined directly with $C^{\alpha }$ as Spin 1 ($%
I_{1}$), 2 ($I_{2}$) and 3 ($I_{3}$), respectively. The methylic hydrogen
nuclei were decoupled using the continuous wave (CW) mode.

The results of the second set of experiment were shown in Fig. 4. The
transitions that connect the $I_{0}^{\alpha }$ and $I_{0}^{\beta }$
manifolds of the observer spin $I_{0}$ are assigned to the computational
basis states of the other spins. The readout on the observer spin $I_{0}$
can tell us which states the computational spins $I_{1}$, $I_{2}$ and $I_{3}$
were in\cite{Madi}. As exhibited in Fig. 4b, only one line corresponding to
the $I_{1}^{\alpha }I_{2}^{\alpha }I_{3}^{\alpha }$ state indicated the
computational spins $I_{1}$, $I_{2}$ and $I_{3}$ was in the $\left|
000\right\rangle $ state. It can been seen from Fig. 4c that, after swapping
Spin 1 and Spin 3, $I_{1}$, $I_{2}$ and $I_{3}$ was in $\left|
000\right\rangle +\left| 100\right\rangle =\left| 0\right\rangle +\left|
4\right\rangle ,$ thus the state periodicity $k=4$ and the state periodicity
before the QFT $r=N/k=2$ was deduced; From Fig. 4d, $I_{1}$, $I_{2}$, $I_{3}$
was in $\left| 000\right\rangle +\left| 010\right\rangle +\left|
100\right\rangle +\left| 110\right\rangle ,$ hence, $k=2$ and deduced $r=4$.
This simple readout method can make one quicker and more intuitive to get
the results and is suitable for any number of qubits. Compared with the
tomography\cite{Chuang} used in the first set of experiments, which is
practically feasible for no more than three qubits, it is simpler and more
efficient. However, its signal-to-noise ratio (SNR) is not very good and it
is more demanding for the sample.

In conclusion, using the NMR techniques, the periodicity of the state can be
extracted from the output state after applying the QFT, which exhibits the
remarkable power of the QFT\ to determine the periodicity. Though the
experimental imperfections cause the reduction of the fidelity, the powerful
ability of quantum computers have been displayed. The QFT can be applied to
many quantum algorithms and other quantum dynamics study. Recently, it has
been used to implement the ordering-finding algorithm\cite{Vandersypen} and
Shor's algorithm in NMR\cite{ShorNMR}.\bigskip

We thank Hanzheng Yuan, and Xu Zhang for help in the course of experiments.

Table 1. Measured NMR parameters for alanine dissolved in $D_{2}O$ on a
Bruker ARX500 spectrometer.

\bigskip

\begin{tabular}{cccccc}
\hline\hline
nuclei & $\nu /Hz$ & $J_{C^{^{\prime }}}/Hz$ & $J_{C^{^{\alpha }}}/Hz$ & $%
J_{C^{\beta }}/Hz$ & $J_{H}/Hz$ \\ \hline
$C^{\prime }$ & $-4320$ &  & $34.94$ & $-1.2$ & $5.5$ \\ \hline
$C^{\alpha }$ & $0$ & $34.94$ &  & $53.81$ & $143.21$ \\ \hline
$C^{\beta }$ & $15793$ & $-1.2$ & $53.81$ &  & $5.1$ \\ \hline
$H$ & $1550$ & $5.5$ & $143.21$ & $5.5$ &  \\ \hline\hline
\end{tabular}

\strut

Table 2. The pulse sequence used to obtain the pseudo-pure state $\rho
_{000}=I_{1}^{\alpha }I_{2}^{\alpha }I_{3}^{\alpha }$ from the thermal
state, $\rho _{eq}=I_{1z}+I_{2z}+I_{3z}.$ The refocusing $\pi $ pulses were
omitted during the evolution $\frac{1}{2J_{12}}$ due to $J_{13}<<J_{12}.$

\bigskip

\begin{tabular}{cl}
\hline\hline
Pulse Sequence for $I_{1}^{\alpha }I_{2}^{\alpha }I_{3}^{\alpha }$ &  \\ 
\hline
\multicolumn{1}{l}{(1) $\left[ \frac{\pi }{3}\right] _{y}^{2}-\left[ \frac{%
5\pi }{12}\right] _{y}^{3}-G_{z}$} &  \\ 
\multicolumn{1}{l}{(2) $\left[ \frac{\pi }{2}\right] _{x}^{1}-\frac{1}{%
2J_{12}}-\left[ -\frac{\pi }{2}\right] _{y}^{1}-G_{z}$} &  \\ 
\multicolumn{1}{l}{(3) $\left[ \frac{\pi }{4}\right] _{x}^{2}-\frac{1}{%
4J_{23}}-\left[ \pi \right] _{x}^{1}-\frac{1}{4J_{23}}-\left[ -\frac{\pi }{4}%
\right] _{y}^{2}-G_{z}$} &  \\ 
\multicolumn{1}{l}{(4) $\left[ \frac{\pi }{4}\right] _{x}^{1}-\frac{1}{%
2J_{12}}-\left[ \frac{\pi }{4}\right] _{y}^{1}-G_{z}$} &  \\ \hline\hline
\end{tabular}

\strut

Table 3 The pulse sequence for creating the labeled pseudo-pure state $\rho
_{000}=I_{0z}I_{1}^{\alpha }I_{2}^{\alpha }I_{3}^{\alpha }$ from the thermal
state, $\rho _{eq}=I_{0z}+I_{1z}+I_{2z}+I_{3z}.$ The refocusing $\pi $
pulses were omitted.

\bigskip

\begin{tabular}{cc}
\hline\hline
Pluse Sequence for $I_{0z}I_{1}^{\alpha }I_{2}^{\alpha }I_{3}^{\alpha }$ & 
\\ \hline
\multicolumn{1}{l}{(1) $\left[ \frac{\pi }{2}\right] _{y}^{1}-\left[ \frac{%
\pi }{2}\right] _{y}^{2}-\left[ \frac{\pi }{2}\right] _{y}^{3}-G_{z}$} & 
\multicolumn{1}{l}{} \\ 
\multicolumn{1}{l}{(2) $\left[ -\frac{\pi }{4}\right] _{y}^{0}-\frac{1}{%
2J_{01}}-\left[ -\frac{\pi }{4}\right] _{x}^{0}-G_{z}$} & \multicolumn{1}{l}{
} \\ 
\multicolumn{1}{l}{(3) $\left[ -\frac{\pi }{4}\right] _{y}^{0}-\frac{1}{%
2J_{02}}-\left[ -\frac{\pi }{4}\right] _{x}^{0}-G_{z}$} & \multicolumn{1}{l}{
} \\ 
\multicolumn{1}{l}{(4) $\left[ -\frac{\pi }{4}\right] _{y}^{0}-\frac{1}{%
2J_{03}}-\left[ -\frac{\pi }{4}\right] _{x}^{0}-G_{z}$} & \multicolumn{1}{l}{
} \\ \hline\hline
\end{tabular}

\newpage

\begin{center}
{\Large Figure Captions}
\end{center}

Fig.1 Efficient circuit for three qubit quantum Fourier transform. $H$
represents the Hadamard gate and conditionality on a spin being in the $%
\left| 1\right\rangle $ state is represented by a filled circle on its time
line. The final symbol inside the box represents the swap $S_{13}.$

Fig.2 Experimental NMR spectra at different stages in the computation. (a)
Experimentally measured thermal equilibrium spectra, acquired after a
read-out pulse $\left( \frac{\pi }{2}\right) _{y}^{i}$ on spin {\it i}. (b)
Experimental spectra for the pseudo-pure state $I_{1}^{\alpha }I_{2}^{\alpha
}I_{3}^{\alpha }.$ As desired, only one line is retained in each multiplet.
(c) and (d) Output spectra of the QFT for the initial state $\left| \psi
\right\rangle =\left| 0\right\rangle +\left| 2\right\rangle +\left|
4\right\rangle +\left| 6\right\rangle $ and $\left| 0\right\rangle +\left|
4\right\rangle ,$ respectively. The positive (negative) lines correspond to
a spin in $\left| 0\right\rangle $ $\left( \left| 1\right\rangle \right) $
before the readout pulse. The positive and negative lines counteract each
other due to a spin in both $\left| 0\right\rangle $ and $\left|
1\right\rangle ,$ so that no line is retained on the spin. (c) and (d) have
been magnified by a factor of two for clarity.

Fig.3 Experimental deviation density matrices: (a) for the pseudo-pure state 
$\left| 000\right\rangle ,$ and for the states obtained after applying the
QFT to the initial state $\left| \psi \right\rangle $ of$,$ (b) $\left|
0\right\rangle +\left| 2\right\rangle +\left| 4\right\rangle +\left|
6\right\rangle ,$ and (c) $\left| 0\right\rangle +\left| 4\right\rangle ,$
respectively. The rows are enumerated in the standard computational basis.
The left and right columns denote the real and imaginary components (in
arbitrary units), respectively.

Fig.4 Experimental NMR spectra for the labeling spin $I_{0}$ at different
stages during the computation. All spectra were recorded by a readout pulse $%
\pi /2$ on $I_{0}.$ (a) Experimentally measured thermal equilibrium spectra.
(b) Experimental spectra for the pseudo-pure state $I_{0z}I_{1}^{\alpha
}I_{2}^{\alpha }I_{3}^{\alpha }$. Only the transition $I_{0x}I_{1}^{\alpha
}I_{2}^{\alpha }I_{3}^{\alpha }$ was presented, corresponding to the pure
state $I_{1}^{\alpha }I_{2}^{\alpha }I_{3}^{\alpha }.$ (c) and (d) Output
spectra of the simplified QFT for the initial state $\left| \psi
\right\rangle =\left| 0\right\rangle +\left| 2\right\rangle +\left|
4\right\rangle +\left| 6\right\rangle $ and $\left| 0\right\rangle +\left|
4\right\rangle ,$ respectively. The spectra of (b), (c) and (d) were added
up 8 times by phase cycling, (c) and (d) were magnified by a factor of two
and four for clarity, respectively.

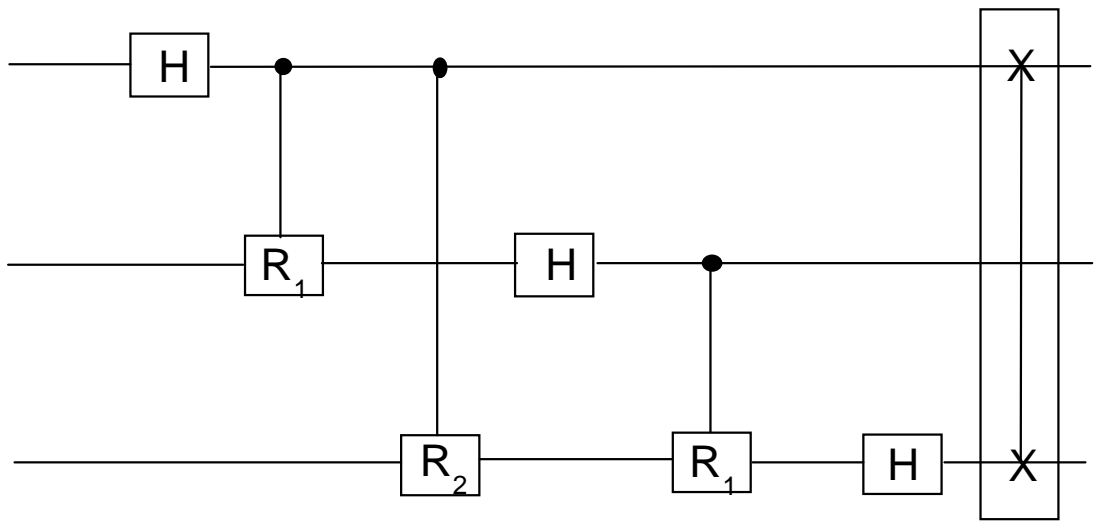
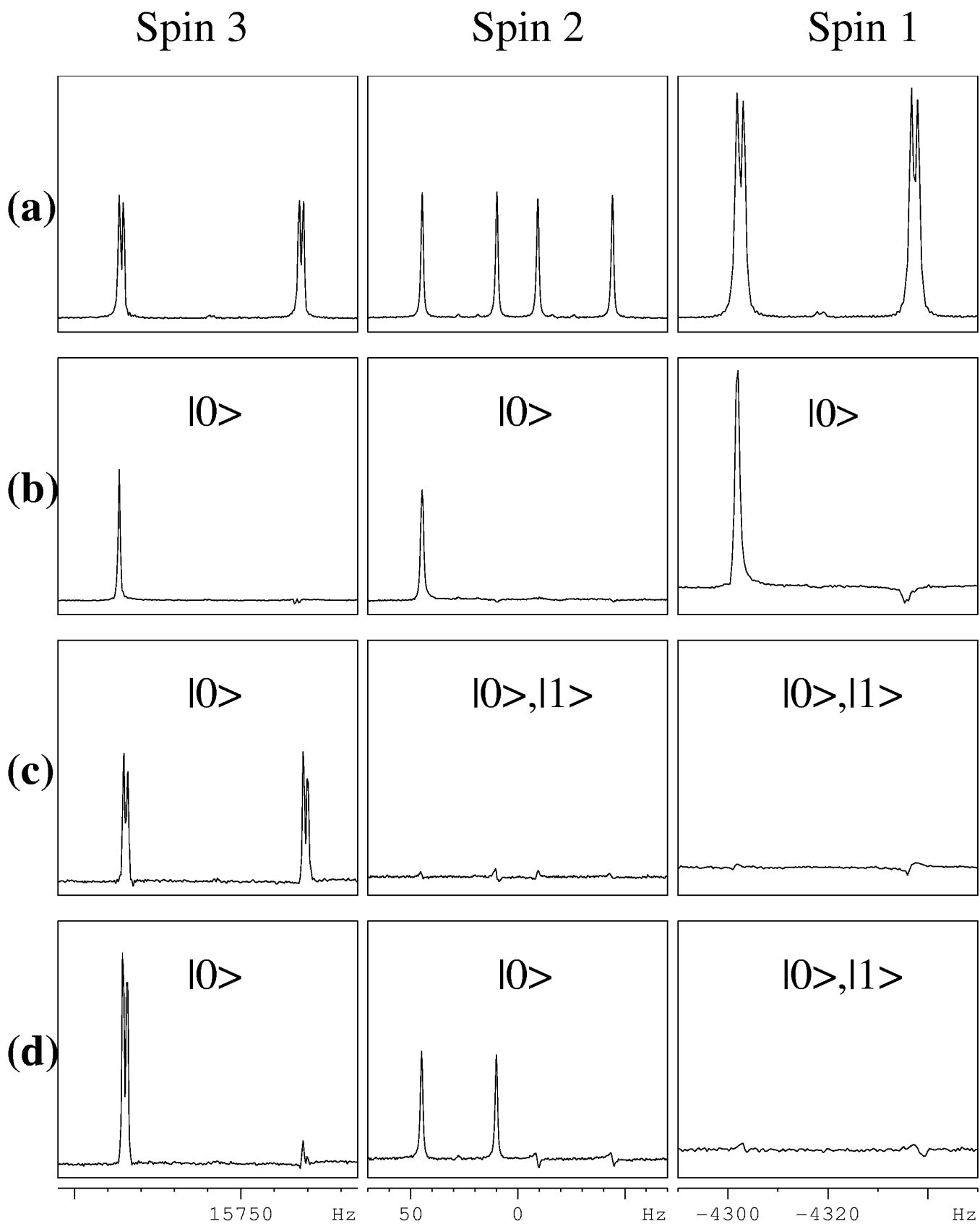
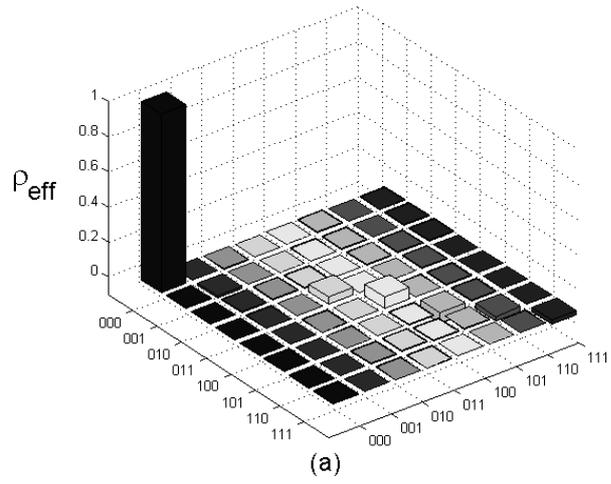
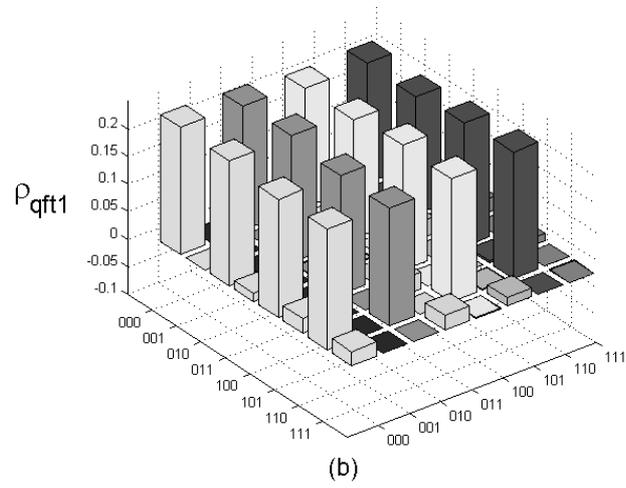
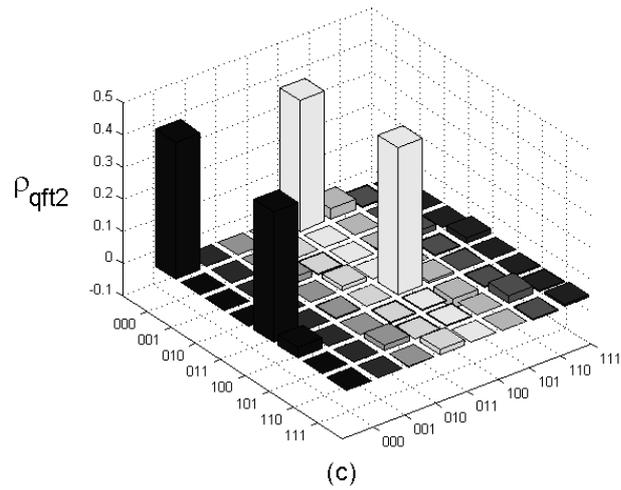
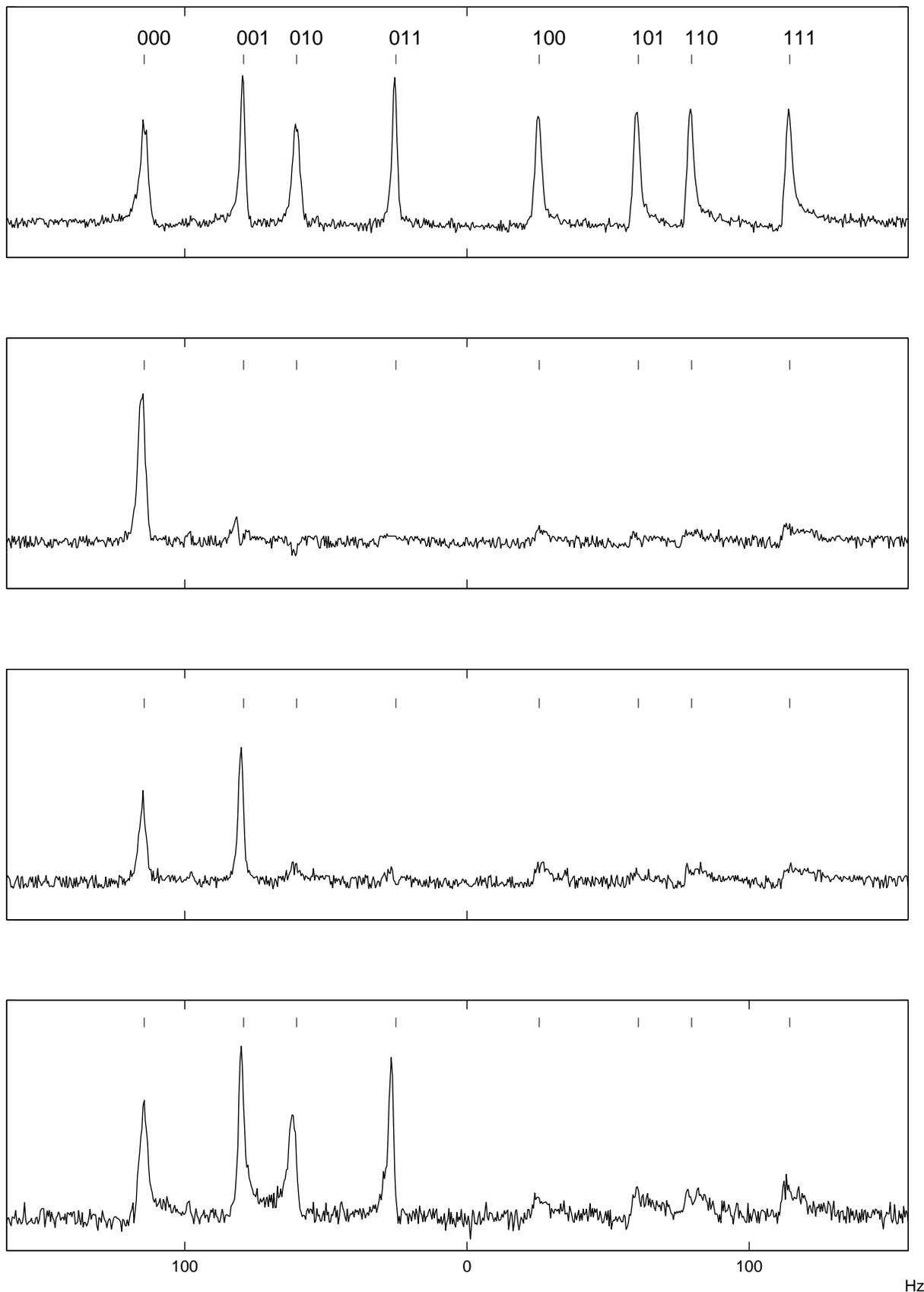
\end{document}